\documentclass[11pt]{amsart}
\usepackage[utf8]{inputenc}
\usepackage{cite,subfigure,fullpage}
\usepackage{url}
\usepackage{wrapfig}
\usepackage{psfrag}
\usepackage{amsmath,amsfonts,amssymb}
\usepackage[dvips]{graphicx}
\usepackage{color}
\usepackage{indentfirst}
\usepackage{epstopdf}
\usepackage{hyperref}

\begin{document}

\title{Mathematical models for COVID-19 pandemic: a comparative analysis\footnote{\MakeLowercase{\MakeUppercase{T}o appear in the "\MakeUppercase{J}ournal of the \MakeUppercase{I}ndian \MakeUppercase{I}nstitute of \MakeUppercase{S}cience," \MakeUppercase{V}olume 100.}}}
\author{Aniruddha Adiga$^1$, \  Devdatt Dubhashi$^2$, \ Bryan Lewis$^1$, \\ Madhav Marathe$^{1,3}$,  \ Srinivasan Venkatramanan$^1$ \ and \ Anil Vullikanti$^{1,3}$\\
$^1$Biocomplexity Institute and Inititiative, University of Virginia \\
$^2$Department of Computer Science and Engineering, Chalmers University \\
$^3$Department of Computer Science, University of Virginia}
\maketitle

\begin{abstract}
COVID-19 pandemic represents an unprecedented global health crisis in the last 100 years.
Its economic, social and health impact continues to grow and is likely to
end up as one of the worst global disasters since the 1918 pandemic and the World Wars. Mathematical models have played an important role in the ongoing crisis; they have been used to inform public policies and have been instrumental in many of the social distancing measures that were instituted worldwide. 

In this article we review some of the important mathematical models used
to support the ongoing planning and response efforts. These models differ
in their use, their mathematical form and their scope. 

\end{abstract}

\maketitle

\section{Introduction}

The ongoing COVID-19 pandemic is the most significant pandemic since the 1918 Influenza pandemic. It has already caused over 21 Million confirmed cases and 758,000 deaths\footnote{The numbers reported are as of August 14, 2020.
See \url{https://coronavirus.jhu.edu/map.html} and
\url{https://nssac.bii.virginia.edu/covid-19/dashboard/} for most up to date
surveillance information}. 
The economic impact is already in trillions of dollars.
As in other pandemics, researchers and public health policy makers are
interested in questions such as\footnote{see \url{https://www.nytimes.com/news-event/coronavirus}},
($i$) How did it start?
($ii$) How is it likely to progress and how can we control it?
($iii$) How can we intervene while balancing public health and economic impact ?
($iv$) Why did some countries do better than other countries thus far into the pandemic?
In particular, models and their projections/forecasts have received unprecedented attention. With a multitude of modeling frameworks, underlying assumptions, available datasets and the region/timeframe being modeled, these projections have varied widely, causing confusion among end-users and consumers. We believe an overview (non-exhaustive) of the current modeling landscape will benefit the readers and also serve as a historical record for future efforts. 

\subsection{Role of models} \ Models have been used by mathematical epidemiologists
to support a broad range of policy questions. Their use during COVID-19
has been widespread. In general, the type and form of models used in epidemiology depends on the phase of the epidemic.  
Before an epidemic, models are used for planning and identifying critical gaps and prepare plans to detect and respond in the event of a pandemic. At the start of a pandemic, policy makers are interested in asking questions such as: ($i$) where and how did the pandemic start, ($ii$) risk of its spread in the region, ($iii$) risk of importation in other regions of the world, ($iv$) basic understanding of the pathogen and its epidemiological characteristics. As the pandemic takes hold researchers begin investigating: ($i$) various intervention and
control strategies; usually pharmaceutical interventions do not work in the event of a pandemic and thus non-pharmaceutical interventions are most appropriate, ($ii$) forecasting the epidemic incidence rate, hospitalization rate and mortality rate, ($iii$) efficiently allocating scarce medical resources to treat the patients and ($iv$) understanding the change in individual and collective behavior and adherence to public policies.
After the pandemic starts to slow down, modelers are interested in developing models related to recovery and long term impacts caused by the pandemic.

As a result comparing models needs to be done with care. When comparing models: one needs to specify: ($a$) the purpose of the model, ($b$) the end user to whom the model is targeted, ($c$) the spatial and temporal resolution of the model, ($d$) and the underlying assumptions and limitations. We illustrate these issues by summarizing a few key methods for \emph{projection and forecasting} of disease outcomes in the US and Sweden.

\medskip
\noindent
\textbf{Organization.} \ The paper is organized as follows. In 
Section~\ref{sec:models} we give preliminary definitions. 
Section~\ref{sec:imperial} discusses US and UK centric models developed by researchers at the Imperial College.
Section~\ref{sec:usa-models} discusses metapopulation models focused on the US that were developed by our group at UVA and the models developed by researchers at Northeastern University. 
Section~\ref{sec:sweden-models} describes models developed Swedish researchers for studying the outbreak in Sweden. In Section~\ref{sec:forecasting} 
we discuss methods developed for forecasting.
Section~\ref{sec:discussion} contains discussion, model limitations and 
concluding remarks. In a companion paper that appears in this special issue, we address certain complementary issues related to pandemic planning and response, including role of data and analytics.

\medskip
\noindent
\textbf{Important note.} \ The primary purpose of the paper
is to highlight some of the salient computational 
models that are currently being used to support COVID-19 pandemic response.
These models, like all models, have their strengths and weaknesses---they have all faced challenges arising from the lack of timely data.
Our goal is \textbf{not} to pick winners and losers among these model; 
each model has been used by policy makers and continues to be used to advice various agencies. Rather, our goal is to introduce to the reader
a range of models that can be used in such situations. A simple model
is no better or worse than a complicated model. The suitability of
a specific model for a given question needs to be evaluated by the 
decision maker and the modeler.

\section{Background: computational methods for epidemiology}
\label{sec:models}
Epidemiological models fall in two broad classes: statistical models that are largely data driven and mechanistic models that are based on underlying theoretical principles developed by scientists on how the disease spreads.

Data-driven models use statistical and machine learning methods to forecast outcomes, such as case counts, mortality and hospital demands. This is a very active area of research, and a broad class of techniques have been developed, including auto-regressive time series methods, Bayesian techniques and deep learning  \cite{adhikari:kdd19, perone:ssrn20, desai:hs19, Reich2019AccuracyOR, funk:epidemics18, murray:ihme}. 
Mechanistic models of disease spread within a population~\cite{Brauer-2008,Ne03, marathe:cacm13,ekmsw-2006} use mechanistic (also referred to as procedural or algorithmic) 
methods to describe the evolution of an epidemic through a population. The most common of these is the SIR type models. Hybrid models that combine mechanistic models with data driven machine learning approaches are also starting to become popular, e.g.,~\cite{Wang2019DEFSIDL}.

\subsection{Mass action compartmental models}
There are a number of models, which are referred to as SIR class of models. These partition a population of $N$ agents into three sets, each corresponding to a disease state, which is one of: susceptible ($S$), infective ($I$) and removed or recovered ($R$). 
The specific model then specifies how susceptible individuals become infectious, and then recover. In its simplest form (referred to as the basic compartmental model)~\cite{Brauer-2008,Ne03, marathe:cacm13}, the population is assumed to be completely mixed. 
Let $S(t)$, $I(t)$ and $R(t)$ denote the number of people who are susceptible, infected and recovered states at time $t$, respectively.
Let $s(t)=S(t)/N$, $i(t)=I(t)/N$ and $r(t)=R(t)/N$; then,
$s(t)+i(t)+r(t)=1$. Then, the SIR model can be described by the following system of ordinary differential equations
\[
\frac{ds}{dt} = \beta si, \qquad \frac{di}{dt} = \beta si - \gamma i, \qquad \frac{dr}{dt} = \gamma i,
\]
where $\beta$ is referred to as the transmission rate, and $\gamma$ is the recovery rate.
A key parameter in such a model is the ``reproductive number'', denoted by $R_0=\beta/\gamma$.
At the start of an epidemic, much of
the public health  effort is focused on estimating $R_0$ from observed  infections~\cite{lipsitch:science03}.

 Mass action compartmental models have been the workhorse for epidemiologists
 and have been widely used for over 100 years. Their strength comes from their simplicity, both analytically and from the standpoint of understanding the
 outcomes. Software systems have been developed to solve such models and a number of associated tools have been built to support analysis using such models.
 
\subsection{Structured metapopulation models}
Although simple and powerful, mass action compartmental models
do not capture the inherent heterogeneity of the underlying populations.
Significant amount of research has been conducted to extend the model, usually in two broad ways.  
The first involves structured metapopulation models---these construct an abstraction of the mixing patterns in the population into $m$ different sub-populations, e.g., age groups and small geographical regions, and attempt to capture the heterogeneity in mixing patterns across subpopulations. In other words, the model has states $S_j(t), I_j(t), R_j(t)$ for each subpopulation $j$. The evolution of a compartment $X_j(t)$ is determined by mixing within and across compartments.
For instance, survey data on mixing across age groups~\cite{Mossong2008SocialCA} has been used to construct age structured metapopulation models~\cite{medlock+optvacc09}. More relevant for our paper are spatial metapopulation models, in which the subpopulations are connected through airline and commuter flow networks~\cite{balcan:pnas2009, srini:ploscb19, gomes:plos-curr14, zhang:pnas17, chinazzi:science20}. 

\medskip
\noindent 
\textbf{Main steps in constructing structured metapopulation models:} This depends on the disease, population and the type of question being studied. The key steps in the development of such models for the spread of diseases over large populations include
\begin{itemize}
\item 
Constructing subpopulations and compartments: the entire population $V$ is partitioned into subpopulations $V_j$, within which the mixing is assumed to be complete. Depending on the disease model, there are $S_j, E_j, I_j, R_j$ compartments corresponding to the subpopulation $V_j$ (and more, depending on the disease)---these represent the number of individuals in $V_j$ in the corresponding state
\item 
Mixing patterns among compartments: state transitions between compartments might depend on the states of individuals within the subpopulations associated with those compartments, as well as those who they come in contact with. For instance, the $S_j\rightarrow E_j$ transition rate might depend on $I_k$ for all the subpopulations who come in contact with individuals in $V_j$. Mobility and behavioral datasets are needed to model such interactions.
\end{itemize}

Such models are very useful at  the early days of the outbreak, when the disease dynamics are
driven to a large extent by mobility---these can be captured more easily within such models,
and there is significant uncertainty in the disease model parameters. They can also model coarser interventions such as reduced mobility between spatial units and reduced mixing rates. However, these models become less useful to model the effect of detailed interventions
(e.g., voluntary home isolation, school closures) on disease spread in and across communities.

\subsection{Agent based network models}
\begin{figure*}
\begin{center}
\includegraphics[width=0.5\linewidth]{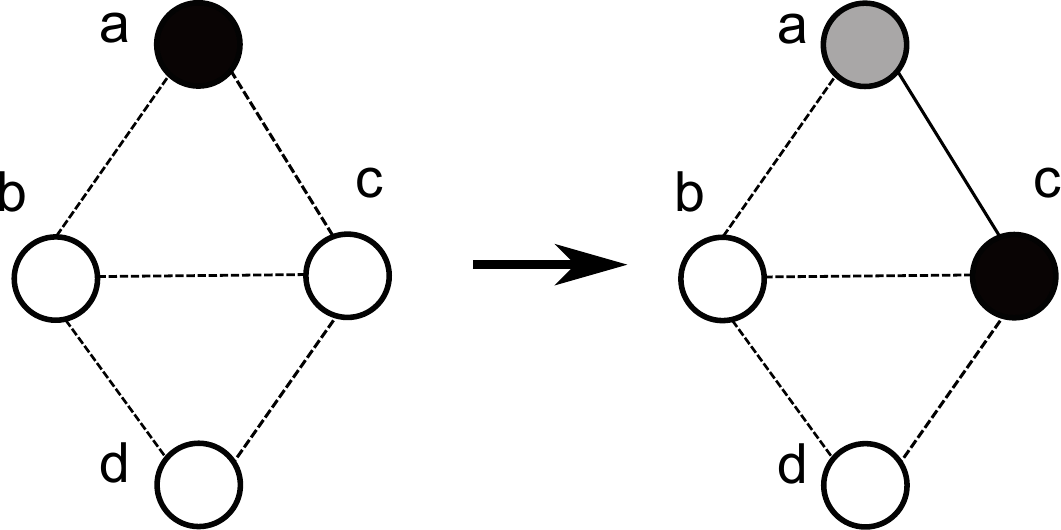}
\caption{The SIR process on a graph. The contact graph $G=(V, E)$ is defined on
a population $V=\{a, b, c, d\}$. The node colors white, black and grey represent
the Susceptible, Infected and Recovered states, respectively. Initially, only node $a$
is infected, and all other nodes are susceptible. A possible outcome at $t=1$
is shown, in which node $c$ becomes infected, while node $a$ recovers. Node $a$ tries
to independently infect both its neighbors $b$ and $c$, but only node $c$ gets
infected--- this is indicated by the solid edge $(a, c)$.
The probability of getting
this outcome is $(1-p(a,b))p(a,c)$.}
\label{fig:network-ex}
\end{center}
\end{figure*}

Agent-based networked models (sometimes just called as agent-based models)  
extend metapopulation models further by explicitly capturing the 
interaction structure of the underlying populations. Often such models are also resolved at the level of single individual entities (animals, humans etc.).
In this class of models,  the epidemic dynamics can be modeled as a diffusion process on a specific undirected contact network $G(V,E)$ 
on a population $V$ -- each edge $e=(u,v)\in E$ implies that
individuals (also referred to as nodes) $u, v\in V$ come into contact\footnote{%
Note that though edge $e$ is
represented as a tuple $(u, v)$, it actually denotes the set $\{u, v\}$, as
is common in graph theory.}
Let $N(v)$ denote the set of neighbors of $v$. 
For instance, in the graph in Figure \ref{fig:network-ex}, we have $V=\{a, b, c, d\}$
and $E=\{(a, b), (a, c), (b, d), (c d)\}$. Node $a$ has $b$ and $c$ as neighbors, so $N(a)=\{b, c\}$.
The SIR model on the graph $G$ is a dynamical process in which
each node is in one of $S$, $I$ or $R$ states.  Infection can
potentially spread from $u$ to $v$ along edge $e=(u,v)$ with a probability of
$\beta(e,t)$ at time instant $t$
after $u$ becomes infected, conditional on node $v$ remaining uninfected
until time $t$--- this is a discrete version of the rate of infection for the
ODE model discussed earlier. 
We let $I(t)$ denote the set of nodes that become
infected at time $t$. 
The (random) subset of edges on which the infections spread represents
a disease outcome, and is referred to as a \emph{dendogram}.
This dynamical system starts with a configuration in which there are one or more nodes
in state {\sf I} and reaches a fixed point in which all nodes are in states {\sf S} or {\sf R}.
Figure \ref{fig:network-ex} shows an example of the SIR model on a network.
\medskip 

\medskip
\noindent 
\textbf{Main steps in setting up an agent based model.} While the specific steps depend on the disease, the population, and the type of question being studied, the general process involves the following steps:
\begin{itemize}
\item 
Construct a network representation $G$: the set $V$ is the population in a region, and is available from different sources, such as Census and Landscan. However, the contact patterns are more difficult to model, as no real data is available on contacts between people at a large scale. Instead, researchers have tried to model activities and mobility, from which contacts can be inferred, based on co-location. Multiple approaches have been developed for this, including random mobility based on statistical models, and very detailed models based on activities in urban regions, which have been estimated through surveys, transportation data, and other sources, e.g.,~\cite{ekmsw-2006,barrett:wsc09,eubank2004modelling, longini05:science,Ferg20}.
\item 
Develop models of within-host disease progression: such models can be represented as finite state probabilistic timed transition models, which are designed in close coordination with biologists, epidemiologists, and parameterized using detailed incidence data (see~\cite{marathe:cacm13} for discussion and additional pointers).
\item 
Develop high performance computer (HPC) simulations to study epidemic dynamics in such models, e.g.,~\cite{barrett2008episimdemics, bisset2009epifast, deodhar2012enhancing, grefenstette2013fred}. Typical public health analyses involve large experimental designs, and the models are stochastic; this necessitates use of such HPC simulations on large computing clusters.
\item 
Incorporate interventions and behavioral changes: interventions include closure of schools and workplaces~\cite{Ferg20,halloran:pnas08} and vaccinations~\cite{eubank2004modelling}, whereas behavioral changes include individual level social distancing, changes in mobility, and use of protective measures.
\end{itemize}

Such a network model captures the interplay between the three components of 
computational epidemiology: ($i$) individual behaviors of agents, 
($ii$) unstructured, heterogeneous multi-scale networks, 
and ($iii$) the dynamical processes on these networks.
It is based on the hypothesis that a better understanding of the
characteristics of the underlying network and individual behavioral
adaptation can give better insights into contagion dynamics and
response strategies.
Although computationally expensive and data intensive, network-based epidemiology
alters the types of questions that can be posed, providing
qualitatively different insights into disease dynamics and public health policies.
It also allows policy makers to formulate
and investigate potentially novel and context specific  interventions.

\subsection{Models for epidemic forecasting}
Like projection approaches, models for epidemic forecasting can be broadly classified into two broad groups: ($i$) statistical and machine learning based data driven models, ($ii$) causal or mechanistic models -- see~\cite{desai:hs19,Reich2019AccuracyOR,nsoesie2014systematic,chretien2014influenza,kandula2019near,tabataba2017framework,brooks2020pancasting}  and the references therein for the current state of the art in this rapidly evolving field.

Statistical methods employ statistical and time-series based methodologies to
learn patterns in historical epidemic data and leverage those patterns for forecasting. Of course the simplest yet useful class is called
{\em method of analogues}. One simply compares the current epidemic with one of the earlier outbreaks and then uses the best match to forecast the
current epidemic. Popular statistical methods for forecasting influenza
like illnesses (that includes COVID-19) 
include e.g. generalized linear models (GLM), autoregressive integrated moving average (ARIMA), and generalized autoregressive moving average (GARMA)~\cite{kandula2019near,LANLforecasts,IHMEcovid2020forecasting}.
Statistical  methods are fast, but they crucially depend on the availability of training data.  Furthermore, since they are purely data driven, they do
not capture the underlying causal mechanisms. As a result epidemic dynamics affected by behavioral adaptations are usually hard to capture.
Artificial neural networks (ANN) have gained increased
prominence in epidemic forecasting due to their self-learning ability without prior knowledge~(See \cite{Wang2019DEFSIDL,wang2020tdefsi,adhikari:kdd19} and the
references therein). 
Such models have used a wide variety of data as surrogates
for producing forecasts. This includes: ($i$) social media data, ($ii$) weather data, ($iii$) incidence curves and ($iv$) demographic data.

Causal models can be used for epidemic forecasting in a natural 
manner~\cite{funk:epidemics18,nsoesie2014systematic,
fadikar2018calibrating,tabataba2017epidemic,chretien2014influenza,yamana2017individual}.
These models, calibrate the internal model parameters using the disease
incidence data seen until a given day and then execute the model forward in time to produce the future time series. Compartmental as well as agent-based models can be used to produce such forecasts. The choice of the models depends on the specific question at hand and the computational and data resource constraints.
One of the key ideas in forecasting is to develop ensemble models -- models that combine forecasts from multiple models~\cite{chakraborty2014forecasting,Reich2019AccuracyOR,yamana2017individual,tabataba2017epidemic}. The idea which originated in the domain of weather forecasting has found methodological advances in the machine learning literature. Ensemble models
typically show better performance than the individual models.

\section{Models from the Imperial College Modeling Group (UK Model)}
\label{sec:imperial}

\noindent
\textbf{Background.} \ The modeling group led by Neil Ferguson
was to our knowledge the first model to study the  impact of 
COVID-19 across two large countries: US and UK, see ~\cite{Ferg20}. 
The basic model was first developed in 2005 -- it was used to inform
policy pertaining to H5N1 pandemic and was one of the three models used
to inform the federal pandemic influenza plan and led to the now well
accepted targeted layered containment (TLC) strategy. It was adapted to COVID-19 as discussed below. The model was widely discussed and covered in the scientific as well as popular press~\cite{adam2020modelling}.
We will refer to this as the IC-model.

\medskip
\noindent
\textbf{Model Structure.} \ The basic model structure consists of 
developing a set of households based on census information for a given country. The structure of the model is largely borrowed from their earlier work, see~\cite{halloran:pnas08,ferguson2006strategies}.
Landscan data was used to spatially distribute the population. Individual members of the household interact with other members
of the household. The data to produce these households is obtained using Census information for these countries. Census data is used to assign age and household sizes. 
Details on the resolution of census data and the dates was not clear. 
Schools, workplaces and random meeting points are then added.
The school data for US was obtained from the National Centre of Educational
Statistics, while for UK schools were assigned randomly based on population density.
 Data on average class sizes and staff-student ratios were used to generate a synthetic population of schools distributed proportional to local population density. Data on the distribution of workplace size was used to generate workplaces with commuting distance data used to locate workplaces appropriately across the population. Individuals are assigned to each of these locations at the start of the simulation. The gravity style kernel is used to decide how far a person can go in terms of attending
work, school or community interaction place. The number of contacts between individuals at school, work and community meeting points are calibrated to produce a given attack rate.

Each individual has an associated disease transmission model. The disease
transmission model parameters are based on data collected when the pandemic
was evolving in Wuhan; see page 4 of \cite{Ferg20}.

Finally, the model also has rich set of interventions. These include:
($i$) case isolation, ($ii$) voluntary home quarantine, ($iii$) Social distancing of those over 70 years, ($iv$) social distancing of the entire population, ($v$) closure of schools and universities; see page 6 \cite{Ferg20}. The code was recently released and is being analyzed. This is important as the interpretation of these interventions can have substantial impact on the outcome.

\medskip
\noindent
\textbf{Model predictions.} \ 
The Imperial college (IC Model) model was one of the first models to evaluate the COVID-19 pandemic using detailed agent-based model. The predictions made by the model were quite dire. The results show that to be able to reduce $R$ to close to 1 or below, a combination of case isolation, social distancing of the entire population and either household quarantine or school and university closure are required.  The model had tremendous impact --
UK and US both decide to start considering complete lock downs -- a
policy that was practically impossible to even talk about earlier in the Western world. The paper came out around the same time that Wuhan epidemic was raging and the epidemic in Italy had taken a turn for the worse. This made the
model results even more critical. 

\medskip
\noindent
\textbf{Strengths and Limitations.} \
IC model was one of the first model by a reputed group to report the
potential impact of COVID-19 with and without interventions. The model
was far more detailed than other models that were published until then.
The authors also took great care parameterizing the model with the best disease transmission data that was available until then. The model also considered a very rich set of interventions and was one of the first to analyze pulsing intervention. On the flip side, the
representation of the underlying social contact network was relatively 
simple. Second, often the details of how interventions were represented were
not clear. Since the publication of their article, the modelers have made their code open and
the research community has witnessed an intense debate on the pros and cons of various modeling assumptions and the resulting 
software system, see~\cite{chawla2020critiqued}. 
We believe that despite certain valid criticisms, overall, the results
represented a significant advance in terms of the 
when the results were put out and the level of details incorporated in the models.

\section{Spatial metapopulation models: Northeastern and UVA models (US Models)}
\label{sec:usa-models}

\noindent
\textbf{Background.}
This approach is an alternative to detailed agent based models, and 
has been used in modeling the spread of multiple diseases, including Influenza~\cite{balcan:pnas2009, srini:ploscb19}, Ebola~\cite{gomes:plos-curr14} and Zika~\cite{zhang:pnas17}. 
It has been adapted for studying the importation risk of COVID-19 
across the world~\cite{chinazzi:science20}.
Structured metapopulation models construct a simple abstraction 
of the mixing patterns in the population, in which the entire region under study is decomposed into fully connected geographical regions,
representing subpopulations, which are connected through airline and commuter flow networks. Thus, they lack the rich detail of agent based models, 
but have fewer parameters, and are therefore, easy to set up and scale to large regions.

\medskip
\noindent
\textbf{Model structure.}
Here, we summarize GLEaM~\cite{balcan:pnas2009} (Northeastern model) and PatchSim~\cite{srini:ploscb19} (UVA model).
GLEaM uses two classes of datasets-- population estimates and mobility.
Population data is used from the ``Gridded Population of the World''~\cite{lloyd2017high}, which gives an estimated population value at a $15\times 15$ minutes of arc (referred to as a ``cell'') over the entire planet. Two different kinds of mobility processes are considered-- airline travel  and commuter flow. The former captures long-distance travel, whereas the latter  captures localized mobility. Airline data is obtained from the International Air Transport Association (IATA)~\cite{IATA}, and
the Official Airline Guide (OAG)~\cite{oag}. There are about 3300 airports world-wide; these are aggregated at the level of urban regions served by multiple airport (e.g., as in London). A Voronoi tessellation is constructed with the resulting airport locations as centers, and the population cells are assigned to these cells, with a 200 mile cutoff from the center.
The commuter flows connect cells at a much smaller spatial scale.
We represent this mobility pattern as a directed graph on the cells, 
and refer to it as the mobility network.

In the basic SEIR model, the subpopulation in each cell $j$ is partitioned into  compartments $S_j, E_j, I_j$ and $R_j$, corresponding to the disease states. For each cell $j$, we define the force of infection 
$\lambda_j$ as the rate at which a susceptible individual in the
subpopulation in cell $j$ becomes infected---this is determined by the interactions the person has with infectious individuals in cell $j$ or any cell $j'$ connected in the mobility network. An individual in the susceptible compartment $S_j$ becomes infected with probability $\lambda_j\Delta t$ and enters the compartment $E_j$, in a time interval $\Delta t$. From this compartment, the individual moves to the $I_j$ and then the $R_j$ compartments, with appropriate probabilities, corresponding to the 
disease model parameters.

The PatchSim~\cite{srini:ploscb19} model has a similar structure, except that it uses administrative boundaries (e.g., counties), instead of
a Voronoi tesselation, which are connected using a mobility network. The mobility network is derived by combining commuter and airline networks, to model time spent per day by individuals of region (patch) $i$ in region (patch) $j$. Since it explicitly captures the level of connectivity through a commuter-like mixing, it is capable of incorporating week-to-week and month-to-month variations in mobility and connectivity. In addition to its capability to run in deterministic or stochastic mode, the open source implementation \cite{NSSACPat38:online} allows fine grained control of disease parameters across space and time. Although the model has a more generic force of infection mode of operation (where patches can be more general than spatial regions), we will mainly summarize the results from the mobility model, which was used for COVID-19 response.

\medskip
\noindent
\textbf{What did the models suggest?}
GLEaM model is being used in a number of COVID-19 related studies and analysis. In ~\cite{kraemer2020effect} the Northeastern University team 
used the model to understand the spread of COVID-19 within China and relative risk of importation of the disease internationally. 
Their analysis suggested that the spread of COVID-19 out of Wuhan into other parts of mainland China was not contained well due to the delays induced by detection and official reporting.
It is hard to interpret the results. The paper suggested that international importation could be contained substantially by strong travel ban. While it might have delayed the onset of cases, the subsequent spread across the world suggest that we were not able to arrest the spread effectively.
The model is also used to provide
weekly projections (see \url{https://covid19.gleamproject.org/}); this site does not appear to be maintained for the most current forecasts
(likely because the team is participating in the CDC forecasting group).

The  PatchSim model is being used to support federal agencies as well
as the state of Virginia. Due to our past experience, we have refrained from providing longer term forecasts, instead focusing on short term projections. The model is used within a \emph{Forecasting via Projection Selection} approach, where a set of counterfactual scenarios are generated based on on-the-ground response efforts and surveillance data, and the best fits are selected based on historical performance. While allowing for future scenarios to be described, they also help provide a reasonable narrative of past trajectories, and retrospective comparisons are used for metrics such as 'cases averted by doing X'. These projections are revised weekly based on stakeholder feedback and surveillance update. Further discussion of how the model is used by the Virginia Department of Health each week can be found at \url{https://www.vdh.virginia.gov/coronavirus/covid-19-data-insights/#model}.

\medskip
\noindent
\textbf{Strength and limitations.}
Structured metapopulation models provide a good tradeoff between the realism/compute of detailed agent-based models and simplicity/speed of mass action compartmental models and need far fewer inputs for modeling, and scalability. This is especially true in the early days of the outbreak, when the disease dynamics are
driven to a large extent by mobility, which can be captured more easily within such models, and there is significant uncertainty in the disease model parameters. However, once the outbreak has spread, it is harder to model detailed interventions (e.g., social distancing), which are much more localized. Further, these are hard to model using a single parameter. Both GLeaM and PatchSim models also faced their share of challenges in projecting case counts due to rapidly evolving pandemic, inadequate testing, a lack of understanding of the number of asymptomatic cases and assessing the compliance levels of the population at large.

\section{Models by  KTH, Umeå and Uppsala researchers (Swedish Models)}
\label{sec:sweden-models}
Sweden was an outlier amongst countries in that it decided to implement public health interventions without a lockdown. Schools and universities were not closed, and restaurants and bars remained open. Swedish citizens
implemented ``work from home'' policies where possible. Moderate social distancing based on individual responsibility and without police enforcement was employed but emphasis was attempted to be placed on shielding the 65+ age group.

\subsection{Simple model}

\medskip
\noindent
\textbf{Background.} \ 
Statistician Tom Britton developed a very simple model with a  focus  on predicting the number of infected over time in Stockholm.

\medskip
\noindent
\textbf{Model structure.} \ 
Britton \cite{Britton2020} used a very simple SIR general epidemic model. It is used to make a coarse grain prediction of the behaviour of the outbreak based on knowing the basic reproduction number $R_0$  and the doubling time $d$ in the initial phase of the epidemic. Calibration to calendar time was done using the observed number of case fatalities, together with estimates of the time between infection to death, and the infection fatality risk. Predictions were made assuming no change of behaviour, as well as for the situation where preventive measures are put in place at one specific time--point. 

\medskip
\noindent
\textbf{Model predictions.} \ One of the controversial predictions from this model was that the number of infections in the Stockholm area would quickly rise towards attaining \emph{herd immunity} within a short period. However, mass testing carried out in Stockholm during June indicated a far smaller percentage of infections.

\medskip
\noindent
\textbf{Strength and Limitations.} \ 
Britton's model was intended as a quick and simple method to estimate and predict an on-going epidemic outbreak both with and without preventive measures put in place. It was intended as a complement to more realistic and detailed modelling. The estimation-prediction methodology is much simpler and straight-forward to implement for this simple model. It is more transparent to see how the few model assumptions affect the results, and it is easy to vary the few parameters to see their effect on predictions so that one could see which parameter-uncertainties have biggest impact on predictions, and which parameter-uncertainties are less influential.

\subsection{Compartmentalized Models I: FHM Model}

\textbf{Background.} \
The Public Health Authority (FHM) of Sweden produced a model
to study the spread of COVID-19 in four regions in Sweden: Dalarna, Sk\aa ne, Stockholm, and V\"astra G\"otaland.\cite{FHM20}.

\medskip
\noindent
\textbf{Model structure.} \ 
It is a standard compartmentalized SEIR model and within each compartment it is homogeneous, so individuals are assumed to have the same characteristics and act in the same way. Data used in the fitting of the model include point prevalences found by PCR-testing in Stockholm at two different time points.

\medskip
\noindent
\textbf{Model predictions.} \ 
The model estimated the number of infected individuals at different time points and the date with the largest number of infectious individuals. It predicted that by July 1, 8.5\% (5.9 – 12.9\%) of the population in Dalarna will have been infected, 4\% (2.4 – 9.9\%) of the population in Skåne will have been infected, 19\% (17.7 – 20.2\%) of the population in Stockholm will have been infected, and 9\% (6.3 – 12.2\%) of the population in Västra Götaland will have been infected.  It was hard to test these predictions because of the great uncertainty in immune response to SARS-CoV-2 -- prevalence of antibodies was surprisingly low but recent studies show that mild cases never seem to develop antibodies against SARS-CoV-2, but only T-cell-mediated immunity \cite{karolinska}.

The model also investigated the effect of increased contacts during the summer that stabilises in autumn. It found that if the contacts in Stockholm and Dalarna increase by less than 60\% in comparison to the contact rate in the beginning of June, the second wave will not exceed the observed first wave.

\medskip
\noindent
\textbf{Strength and limitations.} \ 
The simplicity of the model is a strength in ease of calibration and understanding but it is also a major limitation in view of the well known characteristics of COVID-19: since it is primarily transmitted through droplet infection, the social contact structure in the population is of primary importance for the dynamics of infection. The compartmental model used in this analysis does not account for variation in contacts, where few individuals may have many contacts while the majority have fewer. The model is also not age--stratified, but COVID-19 strikingly affects different age groups differently; e.g., young people seem to get milder infections. In this model, each infected individual has the same infectivity and the same risk of becoming a reported case, regardless of age.  Different age groups normally have varied degrees of contacts and have changed their behaviour differently during the COVID-19 pandemic. This is not captured in the model.

\subsection{Compartmentalized Models II}

\medskip
\noindent
\textbf{Background.} \ 
A group around statistician Joacim Rockl\"ov developed a model to estimate the impact of COVID-19 on the Swedish population at the municipality level, considering demography and
human mobility under various scenarios of mitigation and suppression. They attempted to estimate the
time course of infections, health care needs, and the mortality in relation to the Swedish ICU capacity, as well as the costs of care, and compared alternative policies and counterfactual
scenarios.

\medskip
\noindent
\textbf{Model structure.} \ \cite{Rocklov2020} used a SEIR compartmentalized model with age structured compartments (0-59, 60-79, 80+) susceptibles, infected, in-patient care, ICU and recovered populations based on Swedish population data at the municipal level. It also incorporated inter-municipality travel using a radiation model. Parameters were calibrated based on a combination of values available from international literature and fitting to available outbreak data. The effect of a number of different intervention strategies were considered ranging from no intervention to modest social distancing and finally to imposed isolation of various groups. 

\medskip
\noindent
\textbf{Model predictions.} \ 
The model predicted an estimated death toll of around 40,000 for the strategies based only on social distancing and between 5000 and 8000 for policies imposing stricter isolation. It predicted ICU cases of upto 10,000 without much intervention and upto 6000 with modest social distancing, way above the available capacity of about 500 ICU beds. 

\medskip
\noindent
\textbf{Strength and limitations.} \ 
The model showed a good fit against the reported COVID-19 related deaths in Sweden up to 20th of April, 2020,
However, the predictions of the total deaths and ICU demand turned out to be way off the mark.

\subsection{Agent Based microsimulations}

\noindent
\textbf{Background.} \ 
Finally, \cite{Gardner2020,kamerlin2020managing}used an individual-based model parameterized on Swedish demographics to assess the anticipated spread of COVID-19.

\medskip
\noindent
\textbf{Model structure.} \ 
\cite{Gardner2020} employed the individual agent-based model based on work by Ferguson et al \cite{Ferg20}. Individuals are randomly assigned an age based on Swedish
demographic data and they are also assigned a household. Household size is normally distributed around the average household size in Sweden in 2018, 2.2 people per household. Households were placed on a lattice using high-resolution population data from Landscan and census dara from the Statstics Sweden and each household is additionally allocated to a city based on
the closest city centre by distance and to a county based on city designation. Each individual is placed in a school or workplace at a rate similar to the current participation
in Sweden.

Transmission between individuals occurs through contact at each
individual's workplace or school, within their household, and in their communities. Infectiousness is thus a property dependent on contacts from household members,
school/workplace members and community members with a probability based on household distances. Transmissibility was calibrated against data for the period 21
March – 6 April to reproduce either the doubling time reported using pan-European data or the growth in reported Swedish deaths for that period. Various types of interventions were studied including the policy implemented in Sweden by the public health authorities as well as more aggressive interventions approaching full lockdown.

\medskip
\noindent
\textbf{Model predictions.} \ 
Their prediction was that "under conservative epidemiological parameter estimates, the current Swedish public-health strategy will result in a peak intensive-care load in May that exceeds pre-pandemic capacity by over 40-fold, with a median mortality of 96,000 (95\% CI 52,000 to 183,000)". 

\medskip
\noindent
\textbf{Strength and limitations.} \ 
This model was based on adapting the well known Imperial model discussed in section~\ref{sec:imperial} to Sweden and considered a wide range of intervention strategies. Unfortunately the predictions of the model were woefully off the mark on both counts: the deaths by June 18 are under 5000 and at the peak the ICU infrastructure had at least 20\% unutilized capacity.

\section{Forecasting Models}
\label{sec:forecasting}
Forecasting is of particular interest to policy makers as they attempt to provide actual counts. Since the surveillance systems have relatively stabilized in recent weeks, the development of forecasting models has gained traction and several models are available in the literature. In the US, the Centers for Disease Control and Prevention (CDC) has provided a platform for modelers to share their forecasts which are analyzed and combined in a suitable manner to produce ensemble multi-week forecasts for cumulative/incident deaths, hospitalizations and more recently cases at the national, state, and county level. Probabilistic forecasts are provided by 36 teams as of July 28, 2020 (there were 21 models as of June 24, 2020) and the CDC with the help of \cite{Reichlab} has developed uniform ensemble model for multi-step forecasts \cite{CDChub}.

\subsection{COVID-19 Forecast Hub ensemble model}
It has been observed previously for other infectious diseases that an ensemble of forecasts from multiple models perform better than any individual contributing model \cite{yamana2017individual}. In the context of COVID-19 case count modeling and forecasting, a multitude of models have been developed based on different assumptions that capture specific aspects of the disease dynamics (reproduction number evolution, contact network construction, etc.). The models employed in the CDC Forecast Hub can be broadly classified into three categories, data-driven, hybrid models, and mechanistic models with some of the models being open source. 

\medskip
\noindent
\textbf{Data-driven models:} They do not model the disease dynamics but attempt to find patterns in the available data and combine them appropriately to make short-term forecasts. In such data-driven models it is hard to incorporate interventions directly, hence, the machine is presented with a variety of exogenous data sources such as mobility data, hospital records, etc. with the hope that its effects are captured implicitly. Early iterations of Institute of Health Metrics and Evaluation (IHME) model \cite{IHMEcovid2020forecasting} for death forecasting at state level employed a statistical model that fits a time-varying Gaussian error function to the cumulative death counts and is parameterized to control for maximum death rate, maximum death rate epoch, and growth parameter (with many parameters learnt using data from outbreak in China). The IHME models are undergoing revisions (moving towards the hybrid models) and updated implementable versions are available at \cite{IHMEgithub}. The University of Texas at Austin
COVID-19 Modeling Consortium model \cite{UTwoody2020projections} uses a very similar statistical model as \cite{IHMEcovid2020forecasting} but employs real-time mobility data as additional predictors and also differ in the fitting process. The Carnegie Mellon Delphi Group employs the well known auto-regressive (AR) model that employs lagged version of the case counts and deaths as predictors and determines a sparse set that best describes the observations from it by using LASSO regression \cite{delphi}. \cite{deepGTcovid} is a deep learning model which has been developed along the lines of \cite{adhikari:kdd19} and attempts to learn the dependence between death rate and other available syndromic, demographic, mobility and clinical data.  

\medskip
\noindent
\textbf{Hybrid models}: These methods typically employ statistical techniques to model disease parameters which are then used in epidemiological models to forecast cases. Most statistical models \cite{IHMEcovid2020forecasting, UTwoody2020projections} are evolving to become hybrid models. A model that gained significant interest is the Youyang Gu (YYG) model and uses a machine learning layer over an SEIR model to learn the set of parameters (mortality rate, initial R$_0$,
  post-lockdown R) specific to a region that best fits the region's observed data. The authors (YYG) share the optimal parameters, the SEIR model  and the evaluation scripts with general public for experimentation~\cite{yygeval}. Los Alamos National Lab (LANL) model \cite{LANLforecasts} uses a statistical model to determine how the number of COVID-19 infections changes over time. The second process maps the number of infections to the reported data. The number of deaths are a fraction of the number of new cases obtained and is computed using the observed mortality data.

\medskip
\noindent  
\textbf{Mechanistic models:} GLEaM and JHU models are county-level stochastic SEIR model dynamics. The JHU model incorporates the effectiveness of state-wide intervention policies on social distancing through the R$_0$ parameter. More recently model outputs from UVA's PatchSim model were included as part of a multi-model ensemble (including autoregressive and LSTM components) to forecast weekly confirmed cases. 

\section{Comparative analysis across modeling types}
We end the discussion of the models above by qualitatively comparing
model types. As discussed in the preliminaries, 
at one end of the spectrum are models that are largely data driven: these models range from simple statistical models (various forms
of regression models) to the more complicated deep learning models.
The difference in such model lies in the amount of training data needed,
the computational resources needed and how complicated the mathematical
function one is trying to fit to the observed data. These models are strictly data driven and hence unable to capture the constant behavioral
adaptation at an individual and collective level. On the other end of the
spectrum SEIR, meta-population and agent-based networked models
are based on the underlying procedural representation of the dynamics -- in theory they are able to represent behavioral adaptation endogenously.
But both class of models face immense challenges due to the availability of data as discussed below.

\begin{enumerate}
    \item 
    Agent-based and SEIR models were used in all the three countries in the early part of the outbreak and continue to be used for counter-factual analysis. The primary reason is the lack of 
    surveillance and disease specific data and hence purely data driven models were not easy to use. SEIR models lacked heterogeneity but were simple to program and analyze. Agent-based models were more computationally intensive, required a fair bit of data to instantiate the model but captured the heterogeneity of the underlying countries. 
    By now it has become clear that use of such models for long term forecasting is challenging and likely to lead to mis-leading results. The fundamental reason is adaptive human behavior and lack of data about it.
    \item
    Forecasting on the other hand has seen use of data driven methods as well as causal methods. Short term forecasts have been generally reasonable. Given the intense interest in the pandemic, a lot of data is also becoming available for researchers to use. This helps in validating some of the models further.  Even so, real-time data on
    behavioral adaptation and compliance remains very hard to get and is one of the central modeling challenges.
\end{enumerate}


\section{Models and Policy making}
\label{sec:discussion}
\noindent
\textbf{Were some of the models wrong?} \ 
In a recent opinion piece\footnote{\emph{Indian Express}, July 30, 2020}, Professor Vikram Patel of the Harvard School of Public Health makes a stinging criticism of modelling:
\begin{quote}
    Crowning these scientific disciplines is the field of modelling, for it was its estimates of mountains of dead bodies which fuelled the panic and led to the unprecedented restrictions on public life around the world. None of these early models, however, explicitly acknowledged the huge assumptions that were made,
\end{quote}
A similar article in NY Times recounted the \emph{mistakes} in COVID-19 response in Europe\footnote{NY Times July 20, 2020: \url{https://www.nytimes.com/2020/07/20/world/europe/coronavirus-mistakes-france-uk-italy.html}}; also see~\cite{avery2020policy}.

\medskip
\noindent 
\textbf{Our point of view.} \
It is indeed important to ensure that assumptions underlying mathematical models be made transparent and explicit. But we respectfully 
disagree with Professor Patel's statement: most of the \emph{good} models tried to be very explicit about their assumptions. The mountains of deaths that are being referred to are explicitly calculated when {\bf no} interventions are put in place and are often used as a worst case scenario. Now, one might argue that the authors be explicit and state that this worst case scenario will never occur in practice.
Forecasting dynamics in social systems is inherently challenging: individual
behavior, predictions and epidemic dynamics {\em co-evolve}; this coevolution immediately implies that a dire prediction can lead to extreme change in
individual and collective behavior leading to reduction in the incidence numbers. Would one say forecasts were wrong in such a case or they were
influential in ensuring the worst case never happens? None of this implies that one should not explicitly state the assumption underlying their model.
Of course our experience is that policy makers,
news reporters and common public are looking exactly for such a forecast --
we have been constantly asked "when will peak occur" or "how many people are likely to die". A few possible ways to overcome this tension between the unsatiable appetite for forecasts and the inherent challenges that lie
in doing this accurately, include: 
\begin{itemize}
    \item 
We believe that in general it might not be prudent to provide long term
forecasts for such systems. 
    \item
State the assumptions underlying the models as clearly as possible. Modelers need to be much more disciplined about this. They also need to ensure that the models are transparent and can be reviewed broadly (and expeditiously).
    \item
Accept that the forecasts are provisional and that they will be revised
as new data comes in, society adapts, the virus adapts and we understand the biological impact of the pandemic.
   \item
Improve surveillance systems that would produce data that the models can use more effectively. Even with data, it is very hard to estimate the prevalence of  COVID-19 in society.
\end{itemize}
Communicating scientific findings and risks is an important topical
area in this context, see~\cite{fischhoff2019evaluating,metcalf2020mathematical,adam2020modelling,vaezi2020infodemic}. 

\medskip
\noindent 
\textbf{Use of models for evidence-based policy making.} \
In a new book, \cite{KK20}, \emph{Radical Uncertainty}, economists John Kay and Mervyn King (formerly Governor of the Bank of England) urge caution when using complex models. They argue that models should be valued for the insights they provide but not relied upon to provide accurate forecasts. The so-called "evidence--based policy" comes in for criticism where it relies on models but also supplies a false sense of certainty where none exists, or seeks out the evidence that is desired \emph{ex ante} -- or “cover” -- to justify a policy decision. \emph{"Evidence based policy has become policy based evidence".} 

\medskip
\noindent 
\textbf{Our point of view.} \
The authors make a good point here. But again, everyone, from public
to citizens and reporters clamour for a forecast. We argue that this can
be addressed in two ways:($i$) viewing the problem from the lens 
of control theory so that we forecast only to control the deviation
from the path we want to follow and ($ii$) not insisting on exact numbers but general trends. As Kay and King opine, the value of models, especially in the face of \emph{radical uncertainty}, is more in exploring alternative scenarios resulting from different policies:
\begin{quote}
  a model is useful only if the person using it understands that it does not represent the "the world as it really is" but is a tool for exploring ways in which a decision might or might not go wrong. 
\end{quote}

\medskip
\noindent 
\textbf{Supporting science beyond the pandemic.} \
In his new book \emph{The Rules of Contagion}, Adam Kucharski \cite{K20} draws on lessons from the past. In 2015 and 2016, during the Zika outbreak, researchers planned large-scale  clinical studies and vaccine trials. But these were discontinued as soon as the infection ebbed.  
\begin{quote}
    This is a common frustration in outbreak research; by the time the infections end, fundamental questions about the contagion can remain unanswered. That's why building long term research capacity is essential.
\end{quote}

\medskip
\noindent 
\textbf{Our point of view.} \
The author makes an important point. We hope that today, after witnessing the devastating impacts of the pandemic on the economy and society, the correct lessons will be learnt: sustained investments need to be made in the field to be ready for the impact of the next pandemic.

\textbf{Concluding remarks}
The paper discusses a few important computational models 
developed by researchers in the US, UK and Sweden for COVID-19 pandemic planning and response. The models have been used by 
policy makers and public health officials in their respective 
countries to assess the evolution of the pandemic, design and analyze
control measures and study various what-if scenarios.
As noted, all models faced challenges due to availability of data,
rapidly evolving pandemic and unprecedented control measures put in place.
Despite these challenges, we believe that mathematical models can provide
useful and timely information to the policy makers. On on hand the
modelers need to be transparent in the description of their models, 
clearly state the limitations and carry out detailed sensitivity 
and uncertainty quantification. Having these models reviewed independently
is certainly very helpful. On the other hand, policy makers should be aware
of the fact that using mathematical models 
for pandemic planning, forecast
response rely on a number of assumptions and lack data to over these
assumptions.

\section*{Acknowledgments}
The authors would like to thank members of the Biocomplexity COVID-19 Response Team and Network Systems Science and Advanced Computing (NSSAC) Division for their thoughtful comments and suggestions related to epidemic modeling and response support. We thank members of the Biocomplexity Institute and Initiative, University of Virginia for useful discussion and suggestions. 
This work was partially supported by National Institutes of Health (NIH) Grant 1R01GM109718, NSF BIG DATA Grant IIS-1633028, NSF DIBBS Grant ACI-1443054, 
NSF Grant No.: OAC-1916805, NSF Expeditions in Computing Grant CCF-1918656, CCF-1917819,  NSF RAPID CNS-2028004, NSF RAPID OAC-2027541, US Centers for Disease Control and Prevention 75D30119C05935, DTRA subcontract/ARA S-D00189-15-TO-01-UVA. Any opinions, findings, and conclusions or recommendations expressed in this material are those of the author(s) and do not necessarily reflect the views of the funding agencies.


\bibliographystyle{alpha}
\bibliography{refs}
\end{document}